# KSC2018 우수논문

# 5G 네트워크에서의 무선자원 할당: 신경망을 이용한 접근법

## Radio Resource Allocation in 5G New Radio: A Neural Networks Approach

Madyan Alsenwi, Kitae Kim, and Choong Seon Hong

Department of Computer Science and Engineering, Kyung Hee University, South Korea malsenwi@khu.ac.kr, glideslope@khu.ac.kr, cshong@khu.ac.kr

**요약** 리소스 블록(RB)은 5G 네트워크에서 사용자 장비(UE)에 할당 될 수 있는 최소 주파수-시간 단위이며 한 타임 슬롯 동안의 OFDM(Orthogonal Frequency Division Multiplexing)의 부반송파(subcarrier)로 구성된 채널이다. 이러한 리소스 블록은 15kHz 에서 480kHz 까지 다양한 크기로 나뉠 수 있다. 본 논문에서는 5G 네트워크에서 사용자 장비에 리소스 블록 할당 문제를 GPF(Generalized Proportional Fair) 스케쥴링 공식화 한 후 2 차원 Hopfield 신경망(2D-HNN)으로 모델링하여 해결하는 새로운 방법을 제시한다. 마지막으로 2D-HNN 의 에너지 함수 연구를 통하여 문제를 해결한다. 시뮬레이션 결과 제안된 기법의 효율성을 확인할 수 있었으며 사용자 장비들간 $90\%$ 이상의 공평성(Fairness)를 보장함을 확인하였다.

**키워드:** 5 세대 통신, 자원할당, 신경망, proportional fair.

**Abstract**  The minimum frequency-time unit that can be allocated to User Equipments (UEs) in the fifth generation (5G) cellular networks is a Resource Block (RB). A RB is a channel composed of a set of OFDM subcarriers for a given time slot duration. 5G New Radio (NR) allows for a large number of block shapes ranging from 15 kHz to 480 kHz. In this paper, we address the problem of RBs allocation to UEs. The RBs are allocated at the beginning of each time slot based on the channel state of each UE. The problem is formulated based on the Generalized Proportional Fair (GPF) scheduling. Then, we model the problem as a 2-Dimension Hopfield Neural Networks (2D-HNN). Finally, in an attempt to solve the problem, the energy function of 2D-HNN is investigated. Simulation results show the efficiency of the proposed approach.

## 1. Introduction

The data traffic over cellular networks is continuously increasing. To support this traffic, the Fifth Generation (5G) wireless systems should be able to meet this growing of data traffic. Due to the scarcity of the licensed spectrum, the 5G New Radio (NR) has been extended to support both the 5GHz unlicensed band as well as the millimeter Wave (mmWave) bands. Therefore, 5G NR supports multiple numerologies (i.e., waveform configuration like subframe spacing) and radio frame gets different shapes depending on the type of numerology. The subcarrier spacing is 15 KHz in the low band (less than 3 GHz) outdoor macro network and equals to 30 KHz in outdoor small cell networks. Furthermore, the subcarrier spacing equals 60 KHz in the 5 GHz unlicensed bands and 120 KHz in the mmWave band (28 GHz) [1].

The length of a radio frame is 10 ms and the length of a subframe is always 10 ms regardless of numerology. The main difference in the frame structure is that there is a different number of slots within one subframe. Moreover, the number of symbols within a time slot is different. Therefore, a Resource Block (RB), which is defined as a group of OFDM subcarriers for time slot duration, gets different shapes depending on the numerology. The RB is the smallest frequency-time unit that can be assigned to a User Equipment (UE). The Base station (BS) assigns RBs to UEs as a function of the channel qualities, and traffic demands [1]. Fig. 1 shows the 5G NR numerologies with 15 KHz and 60 KHz sub-carrier spacing.

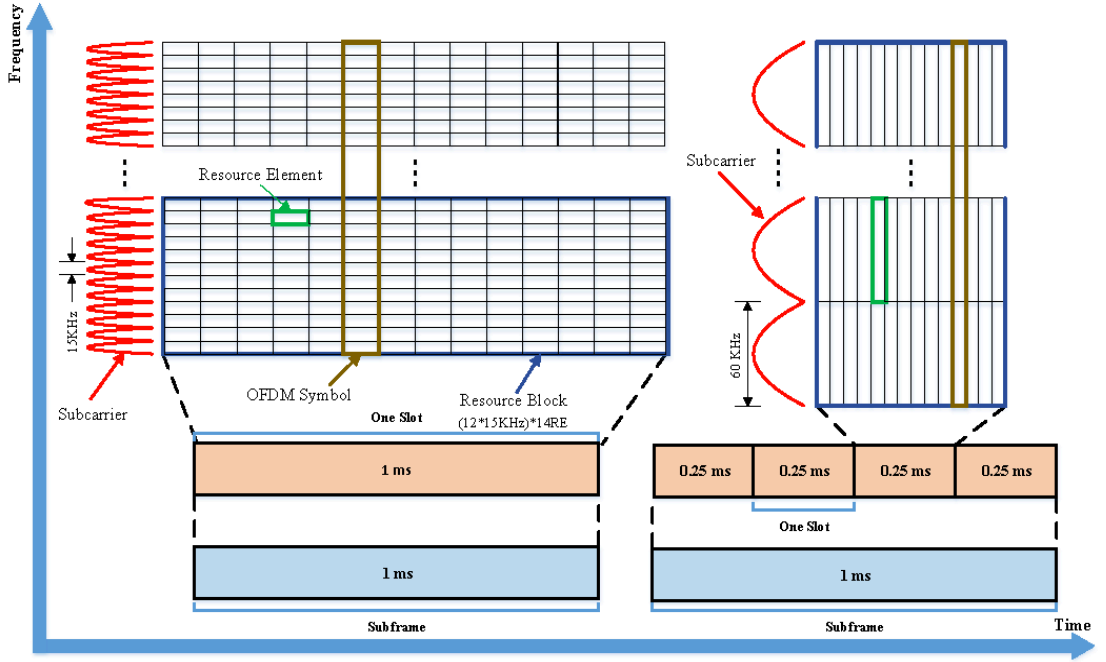

Figure 1: 5G NR Resource Grid for numerologies with sub-carriers of 15 KHz and 60 KHz

Given these different numerologies, the radio resource allocation to different UEs located at different channel conditions and have different traffic demands is crucial to get high network performance. Therefore, we consider the problem of resource allocation to UEs in this work. In literature, two main approaches for the resource allocation problem are proposed: 1) Centralized where there is a central entity (BS) that manages the resource allocation process, 2) Distributed where each user takes its decision autonomously and there is no central entity. A lot of resource allocation mechanisms (centralized and distributed) have been introduced in the literature. Authors in [2] propose a radio resource allocation approach for the green uplink LTE networks. The objective of the proposed algorithm is to allocate the RBs to UEs while maximizing the total throughput considering the single-carrier frequency division multiple access (SC-FDMA) constraints. Authors in [3] introduce an optimization framework to maximize the energy efficiency of the downlink transmission of cellular OFDMA networks while considering discrete power and RBs allocation. The discrete power and discrete RBs are modeled by a single binary variable and a two-stages close-to-optimal semi-definite relaxation (SDR)-based algorithm with Gaussian randomization is proposed to solve the formulated optimization problem. The work in [4] considers the resource slicing to differ-ent traffics in 5G NR named enhanced Mobile Broadband (eMBB) and Ultra-Reliable Low Latency Communication (URLLC). In this work, the RBs are allocated to eMBB users based on a formulation that aims a proportional fair. Furthermore, the resources are allocated to URLLC traffic based on a stochastic formulation that ensures URLLC reliability. In this paper, we model the RBs allocation problem as a 2-Dimension Hopfield Neural Network (2D-HNN). The optimization problem is formulated based on the Generalized Proportional Fair (GPF) formulation and then rewritten in the form of the energy function of 2D-HNN. The decreasing property of HNN energy function is investigated to solve the problem. The main motivation to use the 2D-HNNs is that it can give an online solution due to its ability to process in parallel and thus it can save the computation time.

The remaining of this paper is organized as follows: Section 2 introduces the system model and problem formulation. Section 3 presents the proposed 2D-HNN based approach. Section 4 introduces the performance evaluation. Finally, section 5 concludes the paper.

## 2. System Model and Problem Formulation

We consider the downlink transmissions of a BS with a set of UEs denoted by $\mathcal{U} = \{1, 2, \ldots, U\}$. The BS considers the system bandwidth, which is divided into a set of RBs denoted by $\mathcal{B} = \{1, 2, \ldots, B\}$. The time domain is divided into equally spaced time slots with one millisecond time duration as that in the LTE systems.

The objective is to allocate the RBs to UEs such that the total data rate of all UEs is maximized while ensuring a certain level of fairness between them. The RBs are allocated to UEs at the slot boundary based on their channel states. Therefore, we use channel aware based Generalized Proportional Fair (GPF) scheduling that considers the multi-user diversity [5]. GPF based formulation gives different levels of the trade-off between total data rate and fairness by using different values of the GPF parameter $\alpha$. The data rate and user fairness trade-off optimization problem has to maximize the total data rate of all UEs while maintaining a certain level of long-term fairness. The maximum data rate of a UE u at time slot t can be

approximated based on the Shannon capacity model as follows [6, 7]:

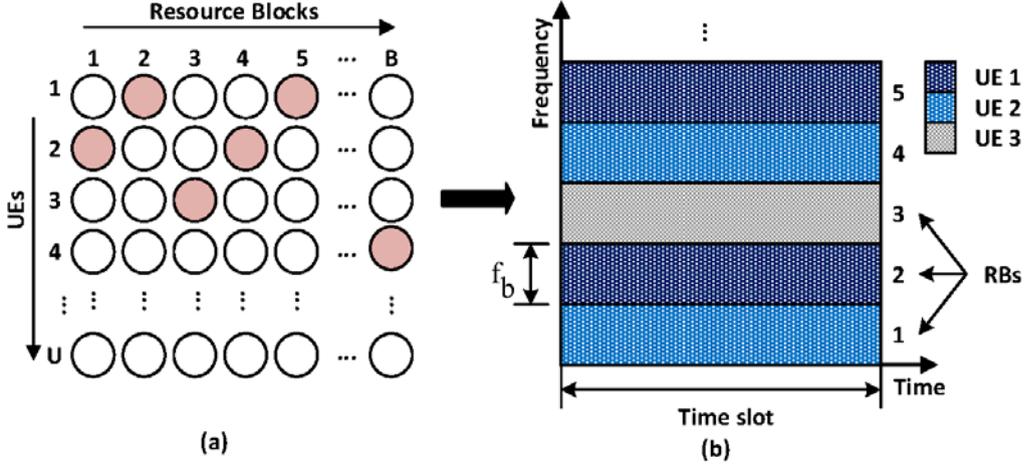

Figure 2: The relation between firing pattern of neural network and RBs allocation

$$r_u(t) = \sum_{b \in \mathcal{B}} f_b x_{u,b}(t) \log_2\left(1 + \frac{p_u h_u(t)}{N_0 F}\right), \quad (1)$$

where $x_{u,b}$ is the RB allocation result, with $x_{u,b} = 1$ means that RB $b$ is allocated to UE $u$ and $x_{u,b} = 0$ means the opposite case, $f_b$ is the bandwidth of RB $b$, $F$ is the total bandwidth, $p_u$ is the transmission power of user $u$, $h_u$ is the channel gain of user $u$, and $N_0$ denotes the noise power.

The average data rate of UE $u$ up to time $t$ can be defined as follows [8]:

$$\bar{R}_u(t) = \epsilon \bar{R}_u(t-1) + (1-\epsilon) r_u(t), \quad (2)$$

where $\epsilon \in [0, 1]$. Therefore, the optimization problem can be formulated as follows:

$$\underset{x}{\text{maximize}} \sum_{u \in \mathcal{U}} \frac{r_u(t)}{[\bar{R}(t)]^\alpha} \quad (3a)$$

$$\text{subject to:} \sum_{b \in \mathcal{B}} x_{u,b} \leq 1, \quad \forall u \in \mathcal{U} \quad (3b)$$

$$x_{u,b} \in \{0,1\}, \quad \forall u \in \mathcal{U}, \forall b \in \mathcal{B}. \quad (3c)$$

The constraint (3b) is to ensure that each RB is allocated to one user only at the same time. The objective is to find the allocation matrix $x$ that maximizes the total data rate of all UEs while ensuring a fair allocation.

## 3. Proposed 2D-Hopfield Neural Networks Based Approach

Artificial Neural Networks (ANNs) are a promising approach to solve optimization problems because it can save computational time due to parallel processing and give an online solution. The ANNs consist of interconnected processing elements called neurons. These neurons work together to solve specific problems. According to its structure, the ANNs are classified into Feed-forward Networks and Feedback Networks or RNNs. Both types have to be configured, one way by training the neural network and letting their weights change according to learning rule. The other way is to set the weights explicitly by using prior knowledge. HNNs, which are a type of RNNs, belong to the non-training model.

In HNNs, the output of each neuron is either '1' or '0' depending on the neuron input (i.e., smaller or larger than its threshold). Every pair of neurons, neuron $i$ and neuron $j$, are connected with the weight $w_{ij}$. In HNNs, The self connections of neurons are set to zero (i.e., $w_{ii} = 0$) and the connections between any two neurons are symmetric (i.e., $w_{ij} = w_{ji}$). The updating rule of neuron i is:

$$v_i(t+1) = \begin{cases} 1, & \text{if } \sum_j w_{ij} v_j(t) \geq \theta_i \\ 0, & Otherwise, \end{cases} \quad (4)$$

where $v_j(t)$ is the state of neuron $j$ at time $t$, and $\theta_i$ is the threshold of neuron $i$. The updating rule of 2D Hopfield Neural Network is given b

$$v_{ij}(t+1) = \begin{cases} 1, & \text{if } \sum_k \sum_l w_{ijkl} v_{kl} \geq \theta_{ij} \\ 0, & Otherwise, \end{cases} \quad (5)$$

where $v_{ij}(t)$ is the state of the neuron $(i,j)$, $w_{ijkl}$ is the connection weight between a neuron $(i,j)$ and neuron $(k,l)$, and $\theta_{ij}$ is the threshold of the neuron $(i,j)$.

HNNs have a value associated with each state of the network called the energy of the network $E(v)$:

$$E(v) = -\frac{1}{2} \sum_i \sum_j w_{ij} v_i v_j + \sum_i \theta_i v_i. \quad (6)$$

The energy function of 2D Hopfield Neural Network can be written as follows:

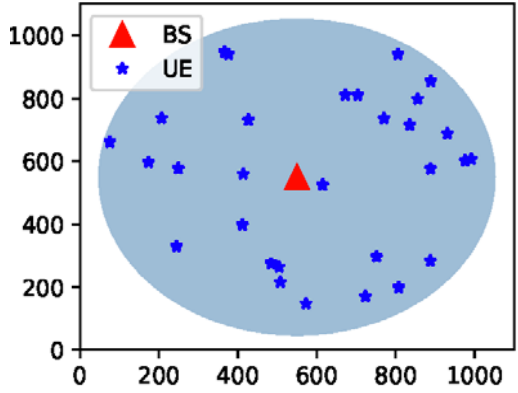

Figure 3: Network Model

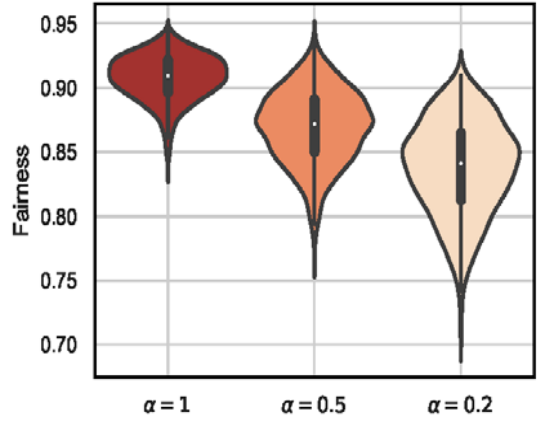

Figure 4: Fairness among UEs for different values of α

$$E(v) = -\frac{1}{2}\sum_i\sum_j\sum_k\sum_l w_{ijkl} v_{ij} v_{kl} + \sum_i\sum_j \theta_{ij} v_{ij}. \quad (7)$$

The value of the energy function decreases when the neurons are updated randomly based on the updating rule and converges to a stable state which is the local minimum of the energy function [9-11]. In this paper, this minimization property is investigated by defining and expressing the objective function in terms of neuron states $v_i$. Then, the connection weights $w_{ij}$ and thresholds $\theta_i$ can be calculated by comparing the energy function with the formulated objective function.

We model the problem of RBs allocation as a 2D-HNN. Here, we consider that there are $B$ neurons for each UE (i.e., number of neurons of each UE equals to the total number of RBs). Therefore, we can express the state of RBs allocation ($x$) of each UE by the firing pattern of the neural network (i.e., Firing neuron $(u,b)$ means that the RB $b$ is assigned to the user $u$ and thus $x_{ub} = 1$). Fig. 2 (a) shows the neural network firing pattern and Fig. 2 (b) shows the corresponding RBs allocation [5].

The objective function of the optimization problem (3) can be written in the same form of HNN energy function as

$$f(x) = -\sum_{u\in\mathcal{U}} \frac{r_u(t)}{[\bar{R}(t)]^\alpha} \quad (8)$$

$$= -\sum_{u\in\mathcal{U}}\sum_{b\in\mathcal{B}} \frac{f_b x_{u,b}}{[\bar{R}(t)]^\alpha} \log_2(1 + \frac{p_u h_u(t)}{N_0 F})$$

$$= -\sum_{u\in\mathcal{U}}\sum_{b\in\mathcal{B}}\sum_{i\in\mathcal{U}}\sum_{j\in\mathcal{B}} \frac{f_b x_{u,b} \delta_{ubij}}{[\bar{R}(t)]^\alpha} \log_2(1 + \frac{p_u h_u(t)}{N_0 F})$$

where $\delta_{ubij}$ is defined as follows:

$$\delta_{ubij} = \begin{cases} 1, & \text{if } u = i \text{ and } b = j \\ 0, & \text{Otherwise}, \end{cases} \quad (9)$$

The connection weights and thresholds can be calculated by comparing $f(x)$ in the above equation with the energy function of 2D-HNN $E(x)$ in equation (7) as follows:

$$w_{ubji} = \frac{f_b x_{u,b} \delta_{ubij}}{[\bar{R}(t)]^\alpha} \log_2(1 + \frac{p_u h_u(t)}{N_0 F}) \quad (10)$$

$$\theta_{ub} = 0, \quad \forall u \in \mathcal{U} \text{ and } b \in \mathcal{B} \quad (11)$$

Each RB can be allocated to one user only at the same time, to satisfy that we modify the updating rule of 2D-HNN in equation (4) to become as follows:

$$x_{ub}(t+1) = \begin{cases} 1, & \text{when } y_{ub}(t+1) = \max[y_{1b}(t+1), \dots, y_{Ub}(t+1)] \\ 0, & \text{Otherwise}, \end{cases} \quad (12)$$

Where

$$y_{ub}(t+1) = \sum_{k\in\mathcal{B}}\sum_{l\in\mathcal{B}} w_{ubkl} x_{kl}(t) - \theta_{ub} \quad (13)$$

## 4. Performance Evaluation

The performance of the proposed approach is evaluated in this section in terms of achieved data rate and fairness. A number of users are distributed randomly in the coverage area of the BS as shown in Fig. 3. The proposed algorithm considers the users at cell edge when allocating resources in order to achieve a fair allocation, i.e. the UEs at cell edge have bad channel state and this impacts their data rate. We consider that 100 RBs are available at each time slot with different width. We evaluate the performance of the proposed approach for different values of the GPF parameter $\alpha$. We calculate the long-term data rate of all UEs and fairness among them for different values of $\alpha$.

Fig. 4 shows the fairness of UEs for different values of $\alpha$. Increasing the value of $\alpha$ leads to higher fairness among the UEs since the RBs allocation algorithm aims to maximize the total data rate at each time slot while considering the average data rate of each user over time. Therefore, the RBs allocation algorithm gives more resources to the users with bad channel conditions, i.e. users at the cell edge. However, setting the value of $\alpha$ to small values decreases the fairness. In this case, the RBs allocation algorithm aims to maximize the total data rate at each time slot without considering the average data rate of each user over time. As a result, the UEs with good channel conditions get more resources. As shown in Fig. 4, the fairness among UEs re-sides between 87% and 95% and the median is almost 92% in the case of $\alpha = 1$. However, the median of the fairness' values decreases to 84% and its values belong between 75% and 90% when setting $\alpha = 0.2$.

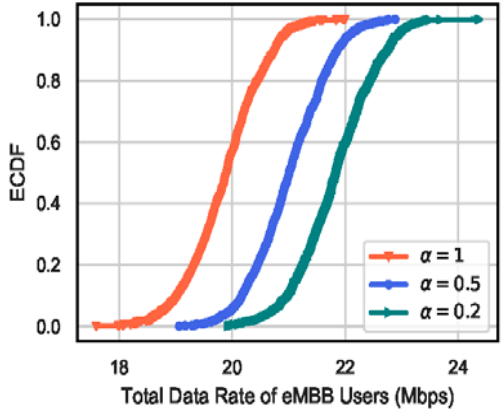

Figure 5: ECDF of the total data rate of all UEs (Mbps) with different values of $\alpha$

Fig. 5 shows the Empirical Cumulative Distribution Function (ECDF) of the total data rate of all UEs for different values of α. As shown in this figure, decreasing the value of $\alpha$ leads to higher total data rate and vice versa. The importance of the average data rate of each user over the time increases when increasing the values of $\alpha$ and this gives a higher probability to the UEs with bad channel conditions to get more resources causing decreasing in the total data rate. Consequently, the RBs allocation algorithm with low values of $\alpha$ allocates more resources to UEs with good channel conditions and this leads to an increase in the total data rate of all UEs. Here, the sum data rate at $\alpha = 0.2$ belongs between 20 Mbps and 24 Mbps with median equals to 22 Mbps. However, the sum data rate belongs between 15Mbps and 22Mbps and its median is 20Mbps when $\alpha = 1$.

## 8. Conclusion

In this paper, we have considered the problem of RBs allocation to UEs in 5G NR. The problem of RBs allocation is formulated based on the GPF. We modeled the problem as a 2D-HNN. Then, the minimization property of energy function of 2D-HNN is investigated to solve it. The results showed that the proposed algorithm gives a fair allocation of the RBs to UEs.

Acknowledgment

This work was supported by the Institute for Information communications Technology Promotion (IITP) grant funded by the Korea government (MSIT) (No.2015-0-00567, Development of Access Technology Agnostic Next-Generation Networking Technology for Wired-Wireless Converged Net-works) *Dr. CS Hong is the corresponding author.

Management, and an Associate Technical Editor of the IEEE Communications Magazine.

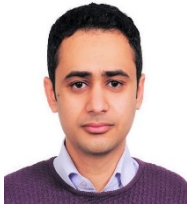

**MADYAN ALSENWI** is current pursuing the Ph.D. degree computer science and engineering with Kyung Hee University, South Ko-rea. Prior to this, he worked as a research assistant under several high impact research projects funded by the Egyptian government. He received the B.E. and MSc degrees in electronics and communications engineering from Cairo University, Egypt, in 2011 and 2016, respectively. His research interests include wireless communications and networking, resource slicing in 5G wireless networks, Ultra Reliable Low Latency Communications (URLLC), UAV-Assisted wireless networks, and machine learning.

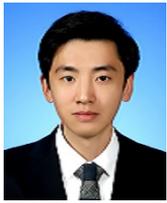

**KITAE KIM** received the B.S and M.S degrees in Computer Science and Engineering at Kyung Hee University, Seoul, Korea, in 2017, 2018 respectively. He is currently pursuing the Ph.D degree in the Computer Science and Engineering at Kyung Hee University, Seoul, Korea. His research interest includes SDN/NFV, wireless network, unmanned aerial vehicle communications, and Machine Learning.

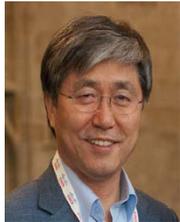

**CHOONG SEON HONG (S'10-M'11-SM'18)** received the B.S. and M.S. degrees in electronic engineering from Kyung Hee University, Seoul, South Korea, in 1983 and 1985, respectively, and the Ph.D. degree from Keio University, Japan, in 1997. In 1988, he joined KT, where he was involved in broadband networks as a Member of Technical Staff. Since 1993, he has been with Keio University. He was with the Telecommunications Network Laboratory, KT, as a Senior Member of Technical Staff and as the Director of the Networking Research Team until 1999. Since 1999, he has been a Professor with the Department of Computer Science and Engineering, Kyung Hee University. His research interests include future Internet, ad hoc networks, network management, and network security. He is a member of the ACM, the IEICE, the IPSJ, the KIISE, the KICS, the KIPS, and the OSIA. He has served as the General Chair, the TPC Chair/Member, or an Organizing Committee Member of international conferences, such as NOMS, IM, APNOMS, E2EMON, CCNC, ADSN, ICPP, DIM, WISA, BcN, TINA, SAINT, and ICOIN. He was an Associate Editor of the IEEE TRANSACTIONS ON NETWORK AND SERVICE MANAGEMENT and the JOURNAL OF COMMUNICATIONS AND NETWORKS. He now serves as an Associate Editor of International Journal of Network